\documentclass[11pt,twoside]{article}

\usepackage{asp2006}
\usepackage{epsf}
\usepackage{psfig}
\usepackage{lscape}

\usepackage{graphics,epsfig}
\usepackage{natbib}

\begin{document}
\title{Astrobiology in the Environments of Main-Sequence Stars: \\
       Effects of Photospheric Radiation}
\author{M. Cuntz and L. Gurdemir}
\affil{Department of Physics, University of Texas at Arlington, Arlington,
TX 76019-0059, USA}
\author{E. F. Guinan}
\affil{Department of Astronomy and Astrophysics, Villanova University,
Villanova, PA 19085, USA}
\author{R. L. Kurucz}
\affil{Harvard-Smithsonian Center for Astrophysics,
Cambridge, MA 02138, USA}

\begin{abstract} 
We explore if carbon-based macromolecules (such as DNA) in the environments of
stars other than the Sun are able to survive the effects of photospheric stellar
radiation, such as UV-C. Therefore, we focus on main-sequence stars of spectral
types F, G, K, and M.  Emphasis is placed on investigating the radiative environment
in the stellar habitable zones.  Stellar habitable zones are relevant to astrobiology
because they constitute circumstellar regions in which a planet of suitable size
can maintain surface temperatures for water to exist in fluid form, thus
increasing the likelihood of Earth-type life.
\end{abstract}



\section{Theoretical approach}

The centerpiece of all life on Earth is carbon-based biochemistry. It has
repeatedly been surmised that biochemistry based on carbon may also play a
pivotal role in extraterrestrial life forms, if existent \citep[e.g.,][]{ben07}.
This is due to the pronounced advantages of carbon, especially compared to its
closest competitor (i.e., silicon), which include: its relatively high
abundance, its bonding properties, and its ability to form very large molecules
as it can combine with hydrogen and other molecules as, e.g., nitrogen and oxygen
in a very large number of ways \citep{gol02}.

In the following, we explore the relative damage to carbon-based macromolecules
in the environments of stars other than the Sun using DNA as a proxy.
We focus on the effects of photospheric radiation from main-sequence stars,
encompassing the range between F0 and M0.  Our models consist of the following
components:

\begin{enumerate}

\item
The radiative effects on DNA are considered by applying a DNA action spectrum
\citep{hor95}.  It shows that the damage is strongly wavelength-dependent,
increasing by more than seven orders of magnitude between 400 and 200~nm.
The different regimes are commonly referred to as UV-A, UV-B, and UV-C.

\item
The planets are assumed to be located in the stellar habitable zone (HZ).
Following the concepts by \cite{kas93} and \cite{und03}, we distinguish
between the conservative and generalized HZ (see Fig. 2).  The inner and
outer edge of the conservative HZ are given by the onset of water loss and
CO$_2$ condensation, respectively, whereas the inner and outer edge of the
generalized HZ are given by the runaway greenhouse effect and the breakdown
of greenhouse heating, respectively, needed to permit the existence
of fluid water on the planetary surface.

\begin{figure*}
\centering
\epsfig{file=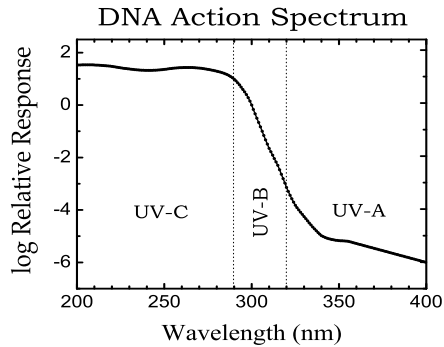,width=0.5\linewidth,height=0.39\linewidth}
\caption{DNA action spectrum following \cite{hor95}.
}
\end{figure*}

\begin{figure*}
\centering
\epsfig{file=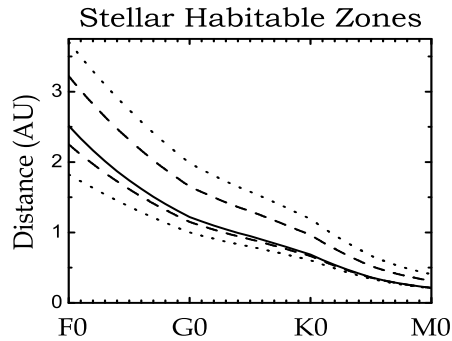,width=0.5\linewidth,height=0.39\linewidth}
\caption{Stellar habitable zones for main-sequence stars of different
spectral types.  The dashed lines indicate the conservative HZ, whereas
the dotted lines indicate the generalized HZ.  The solid line denotes
the Earth-equivalent position.
}
\end{figure*}

\item
Stellar photospheric radiation is represented by using realistic spectra,
which take into account millions or hundred of millions of lines for
atoms and molecules \citep[][and related publications]{cas04}.
Clearly, significant differences emerge between the different spectral types,
both concerning the total amount of radiation and their spectral distribution.

\item
We also consider the effects of attenuation by a planetary atmosphere.  The
following cases are considered: Earth as today, Earth 3.5 Gyr ago, and no
atmosphere at all \citep{coc02}.  For general discussions see, e.g.,
\cite{gui02} and \cite{gui03}.

\end{enumerate}

Our results are presented in Figs. 3 and 4.  They show the relative
damage to DNA due to photospheric radiation from stars between spectral
type F0 and M0.  The results are normalized to today's Earth, placed at
1 AU from a star of spectral-type G2V.  We also considered
planets at the inner and outer edge of either the conservative or
generalized HZ as well as planets of different atmospheric
attentuation.

\begin{figure*}
\centering
\epsfig{file=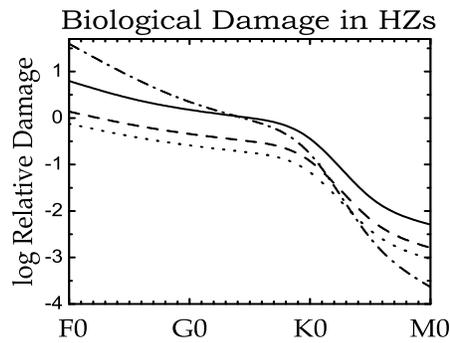,width=0.5\linewidth,height=0.39\linewidth}
\caption{Biological damage to DNA for a planet
at an Earth-equivalent position without an atmosphere
(solid line), an atmosphere akin to Earth 3.5 Gyr ago (dashed line)
and an atmosphere akin to Earth today (dotted line).  The dash-dotted
line refers to a planet without an atmosphere at 1~AU.
}
\end{figure*}

\begin{figure*}
\centering
\epsfig{file=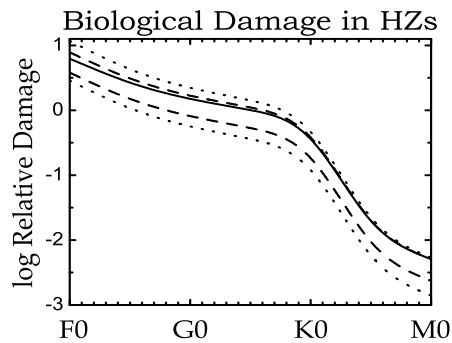,width=0.5\linewidth,height=0.39\linewidth}
\caption{Biological damage to DNA for a planet (no atmosphere)
at an Earth-equivalent position (solid line), at the limits of the
conservative HZ (dashed lines) and at the limits of the generalized HZ
(dotted lines).
}
\end{figure*}

\newpage

\section{Conclusions}

Based on our studies we arrive at the following conclusions:

\begin{enumerate}

\item
All main-sequence stars of spectral type F to M have the potential of
damaging DNA due to UV radiation.  The amount of damage strongly depends
on the stellar spectral type, the type of the planetary atmosphere and
the position of the planet in the habitable zone (HZ).  Our results
constitute a quantitative update and improvement of previous work by
\cite{coc99}.

\item
The damage to DNA for a planet in the HZ around an F-star (Earth-equivalent
distance) due to photospheric radiation is significantly higher (factor 5) compared
to planet Earth around the Sun, which in turn is significantly higher than for an
Earth-equivalent planet around an M-star (factor 180).
Small modifications of this picture occur for different planetary positions
inside their respective HZs.

\item
Regarding the cases studied, we found that the damage is most severe in
the case of no atmosphere at all, somewhat less severe for an atmosphere
corresponding to Earth 3.5 Gyr ago, and least severe for an atmosphere like
Earth today.

\end{enumerate}

Our results are of general interest for the future search for planets
in stellar HZs, the chief goal of future NASA search missions
\citep[e.g.,][]{tur03}.  Our results also reinforce the notion
that habitability may at least in principle be possible around M-type stars, as
previously discussed by \cite{tar07}.  Note, however, that a more detailed
analysis also requires the consideration of chromospheric UV radiation,
especially stellar flares \citep[e.g.,][]{rob05}.


\end{document}